\documentclass[12pt]{article}

\usepackage{epsfig}
\usepackage{amssymb}

\def\be{\begin{equation}}
\def\ee{\end{equation}}
\def\ba{\begin{array}{c}}
\def\ea{\end{array}}

\def\ben{$$}
\def\een{$$}

\newcommand{\kkt}{\kt\!\kt}
\newcommand{\pbr}{\prec\!}
\newcommand{\pkt}{\!\!\succ\,\,}
\newcommand{\kt}{\rangle}

\def\bea{\begin{eqnarray}}
\def\eea{\end{eqnarray}}

\def\beax{\begin{eqnarray*}}
\def\eeax{\end{eqnarray*}}

\begin{document}


\vspace{.35cm}

\begin{center}

{\Large

Quantum control and the challenge of non-Hermitian model-building

}

\vspace{10mm}

\textbf{Miloslav Znojil}

\vspace{3mm} Nuclear Physics Institute ASCR,

250 68 \v{R}e\v{z}, Czech Republic

{e-mail: znojil@ujf.cas.cz}


\end{center}



\section*{Abstract}

In a way inspired by the brief 2002 note ``The challenge of
nonhermitian structures in physics'' by Ramirez and Mielnik (with
the text most easily available via arXiv: quant-ph/0211048) the
situation in the theory is briefly summarized here as it looks
twelve years later. Our text has three parts. In the first one we
briefly mention the pre-history (dating back to the Freeman Dyson's
proposal of the non-Hermitian-Hamiltonian method in 1956 and to its
subsequent successful ``interacting boson model'' applications in
nuclear physics) and, first of all, the amazing recent progress
reached, in the stationary case, using, in essence, an inversion of
the Dyson's approach. The impact on the latter idea upon abstract
quantum physics is sampled, first of all, by the reference to papers
by Bender et al (who made the non-Hermitian model-building popular
under the nickname of parity-times-time-reflection-symmetric {\it
alias} PT-symmetric quantum mechanics) and by Mostafazadeh (who
reinterpreted PT-symmetry as P-pseudo-Hermiticity). In the second
part of our review the emphasis is shifted to the newest,
non-stationary upgrade of the formalism which we proposed in the
year 2009 and which is characterized by the simultaneous
participation of a triplet of Hilbert spaces ${\cal H}$ in the
representation of a single quantum system. In the third part of the
review we finally emphasize that the majority of applications of our
three-Hilbert-space (THS) recipe is still ahead of us because the
enhancement of the flexibility is necessarily accompanied by an
enhancement of the technical difficulties. An escape out of the
technical trap is proposed to be sought in a restriction of
attention to quantum models living in finite-dimensional Hilbert
spaces ${\cal H}$. As long as the use of such spaces is so typical
for the quantum-control considerations, we conclude with conjecture
that the THS formalism should start searching for implementations in
the field of quantum control.

%
%
%
%
%

\newpage

\section{Introduction and summary.}

\subsection{Schr\"{o}dinger equation and quantum control.}

In accord with the standard textbooks on quantum theory \cite{[94]}
the evolution of a pure state of a closed quantum system is most
comfortably determined by solving Schr\"{o}dinger equation
 \begin{equation}
 \label{schrodinger}
 i\hbar\frac{\partial}{\partial t}|\psi(t)\rangle^{(P)} =
 \mathfrak{h}(t)|\psi(t)\rangle^{(P)}
 \end{equation}
in which one assumes that the states of the system in question are
represented by elements $|\psi(t)\rangle^{(P)}$ of a properly
selected physical Hilbert space ${\cal H}^{(P)}$ and in which the
generator of evolution (called Hamiltonian) is self-adjoint,
$\mathfrak{h}(t)=\mathfrak{h}^\dagger(t)$.

Using the notation conventions of Ref.~\cite{SIGMA} and abbreviating
$|\psi(t)\rangle^{(P)}\,\equiv\,|\psi(t)\pkt$ one traditionally
assumes that at $t=0$ the system is prepared in an initial state
$|\psi(0)\pkt=|\psi_{i}\pkt$ and that it is detected, after some
time $T>0$, in a final state $|\psi(T)\pkt=|\psi_{f} \pkt$. Thus, in
the most conventional approach one knows $\mathfrak{h}(t)$ and
constructs the final state $|\psi_{f}\pkt \in {\cal H}^{(P)}$.

An entirely different task is typical for the so called
quantum-control (QC) setup. In its most elementary specification
(often called bilinear model -- see, e.g., review paper \cite{Dong}
for details) one is given just a desirable final target state
$|\psi(T)\pkt=|\psi_{f}\pkt$. For the purpose, one has to select a
suitable ``realization Hamiltonian''
$\mathfrak{h}(t)=\mathfrak{h}^\dagger(t)$ and specify the necessary
``realization time'' $T>0$.

For technical reasons people often restrict their attention to
finite-dimensional quantum systems living in an $N$-dimensional
complex Hilbert spaces ${\cal H}^{(P)} = \ell^{(N)}$ in which the
admissible self-adjoint QC Hamiltonians have the form of a
superposition
 \be
 \mathfrak{h}(t)=\mathfrak{h}_{0} +
 \sum_{k=1}^Ku_{k}(t)\mathfrak{h}_{k}\,.
 \label{postulate}
 \ee
The $(K+1)-$plet of auxiliary operators $\{\mathfrak{h}_{0},
\mathfrak{h}_{1}, \dots, \mathfrak{h}_{K}\}$ is assumed
time-independent. Moreover, this multiplet of operators is often
chosen as a set of generators of a Lie algebra $\mathcal{L}_{0}$
such that the desired evolution of the system towards a given target
state may be proved to exist (one speaks about a ``controllability''
\cite{[25]}). In such a setting the target $|\psi_{f}\pkt$ is to be
reached solely via the selection of the real coefficients
$u_{k}(t)\in \mathbb{R}$ called {\em control functions}.

\subsection{The plan and summary of the paper.}

In our present paper we intend to expose the standard,
above-outlined formulation of the quantum control problem to a
modification. It will be inspired by the recent developments in
quantum theory in which one complements the {\em ad hoc} choice of
dynamics (i.e., of the Hamiltonian) by the possibility of an {\em
independent} alteration of the Hilbert space itself.

The essence of the latter developments will be explained in sections
\ref{jednasec}, \ref{jednasecbe} and \ref{dvasec}. Firstly, in
section \ref{jednasec} we shall follow, for pedagogical reasons,
older reviews \cite{Carl,ali} and introduce the amendment
$\mathfrak{h} \to G \neq G^\dagger$ of the generator of evolution in
its simplified, time-independent version with $\mathfrak{h}\neq
\mathfrak{h}(t)$ replaced by $ G \neq  G(t)$ in Schr\"{o}dinger
Eq.~(\ref{schrodinger}). Marginally, let us add that from the
present perspective, sections \ref{jednasec} and \ref{jednasecbe}
should be read as a mere contextual introduction. They offer a
review of recent updates of quantum theory which  became widely
known as ${\cal PT}-$symmetric Quantum Mechanics (PTSQM, cf. the
Bender's review paper \cite{Carl}) or, in a slightly more general
form, as Pseudo-Hermitian representation of Quantum Mechanics
(PHRQM, cf. the Mostafazadeh's review paper \cite{ali}). As long as
in both of these approaches the operators of observables must remain
stationary (or, at best, quasi-stationary \cite{PLB}) none of these
formalism is directly applicable in the QC context.

Subsequently, section \ref{dvasec} will outline the upgraded and
generalized (a.k.a. ``three-Hilbert-space'', THS) formalism of
Refs.~\cite{SIGMA,timedep}. Our discussion will cover the case in
which the manifest time dependence of $\mathfrak{h}=
\mathfrak{h}(t)$ and of $ G =  G(t)$ is permitted. Although some of
the preceding ideas remain unchanged, it will be necessary to shift
the emphasis. Indeed, only a change of perspective will enable us to
address the QC-related conceptual questions. In this sense, the next
section \ref{dvasecbe} should be read as a more technical addendum
reviewing a few aspects of necessary mathematics. The key message is
that the most general time-dependent Hamiltonian-like operators may
still be required to generate the standard {\em unitary} evolution
of a given quantum system in time. We shall also explain why, in
contrast to the PTSQM or PHRQM scenarios, the spectra of our present
Hamiltonian-like generators $G(t) \neq G^\dagger(t)$ are, in
general, complex.

In the key part and climax of our message in section \ref{trisec} we
shall return to the problems of quantum control, outlining briefly
the possible use of the whole THS machinery for an enhancement of
the flexibility and efficiency of the specific QC tasks.
Preliminarily, our proposal may be summarized as opening a new
approach to quantum control in which one extends the model-building
freedom via a transfer of Schr\"{o}dinger equation from its
representation (\ref{schrodinger}) in the ``primary'' Hilbert space
${\cal H}^{(P)}$ (which is assumed to appear, for any reason,
unfriendly) to some of its alternative though, by assumption,
equivalent and technically friendlier forms.

A few complementary comments on such a possibility will be finally
formulated in our last section \ref{point}. We shall emphasize that
the THS-representation-mediated introduction of the manifestly
time-dependent non-Hermitian generators of evolution $G(t)$ is in
fact necessary in the QC context. We believe that our considerations
will offer a sufficiently strong encouragement for a more concrete
model-building activity in the nearest future.

\section{Time-independent non-Hermitian Hamiltonians in Quantum
Mechanics.
 \label{jednasec}}

\subsection{Modified Schr\"{o}dinger equation.}

It is well known \cite{[94]} that in principle, the constructive
solution of Schr\"{o}dinger Eq.~(\ref{schrodinger}) is particularly
straightforward for Hamiltonians which are Hermitian, diagonalized
and {\em not} time-dependent. Even in these cases, there exist
quantum systems (like, for example, heavy atomic nuclei) for which
even the brute-force numerical diagonalization of a given Hermitian
$\mathfrak{h}\neq \mathfrak{h}(t)$ yields, typically, very poorly
convergent results.

In the year 1956, one of the most unexpected ways out of similar
difficulties has been proposed by Dyson \cite{Dyson}. He proposed a
reparametrization $|\psi(t)\rangle^{(P)}
=\Omega\,|\psi(t)\rangle^{(F)} $ of wave functions in which a
``friendlier'' ket $|\psi(t)\rangle^{(F)} $ was assumed to belong to
a  ``friendlier'' Hilbert space ${\cal H}^{(F)}$. Moreover, the
time-independent mapping $\Omega$ was chosen, in contrast to common
practice, {\em non-unitary}, yielding a nontrivial operator product
$\Omega^\dagger\Omega \neq I$. In this way, the insertion in
Eq.~(\ref{schrodinger}) led to a potentially friendlier
Schr\"{o}dinger equation defined in the new Hilbert space,
 \begin{equation}
 \label{Fschrodinger}
 i\hbar\frac{\partial}{\partial t}|\psi(t)\rangle^{(F)} =
 H\,|\psi(t)\rangle^{(F)}\,,
 \ \ \ \ \ \ H = \Omega^{-1}\mathfrak{h}\Omega \neq H(t)\,.
 \end{equation}
Naturally, in the numerical setting the Dyson's trick and
Hilbert-space invertible mapping $\Omega: {\cal H}^{(F)} \to {\cal
H}^{(P)}$ only made sense if it led to an accelerated convergence
but its enormous success may be found confirmed, e.g., in the recent
nuclear-physics-devoted review paper \cite{Geyer}.

A not entirely pleasant consequence of the non-Hermiticity of the
Dyson's mapping may be seen in the emergence of a manifest
non-Hermiticity of $H$. Indeed, in the new language the old
Hermiticity rule reads
 \be
 \mathfrak{h}^\dagger
 = [\Omega^{-1}]^\dagger
 \,H^\dagger\, \Omega^\dagger
 =\Omega \,H\, \Omega^{-1}
 \ee
and may be re-written in a more compact form
 \be
 H^\dagger\, \Theta
 = \Theta \,H\,,
 \ \ \ \ \  \Theta= \Omega^\dagger\,\Omega \neq I\,.
 \label{Dieudonne}
 \ee
Thus, the new Hamiltonian may only be declared ``quasi-Hermitian''
\cite{Dieudonne}.

\subsection{The coexistence and mutual relations of
the triplet of simultaneous representation Hilbert spaces.}

A new life of the same old trick has been conceived in 1998 when
Bender with Boettcher \cite{BB} proposed the use of certain
non-Hermitian $H$ with real spectrum in the role of a standard
quantum energy observable. Subsequently, the consistent PTSQM
formalism (with its physics-inspired emphasis on the additional
feature of parity-times-time-reversal symmetry) has been born, in
its final form, in the year 2004 \cite{Carl,erratum}. In parallel,
also the more general, less restrictive PHRQM version of the
formalism as already known to nuclear physicists before 1992
\cite{Geyer} was given new life and popularity by Mostafazadeh (see
his numerous publications and/or their summary in his comprehensive
review paper \cite{ali}).

For our present purposes the PHRQM relations between the
above-mentioned P- and F-superscripted Hilbert spaces (and between
these two spaces and the third, S-superscripted space which only
differs from the F-space by the use of the metric-mediated, i.e.,
$\Theta-$mediated inner product) may be summarized using the
following diagram,
 \ben
  \vspace{-1cm}
  \ba
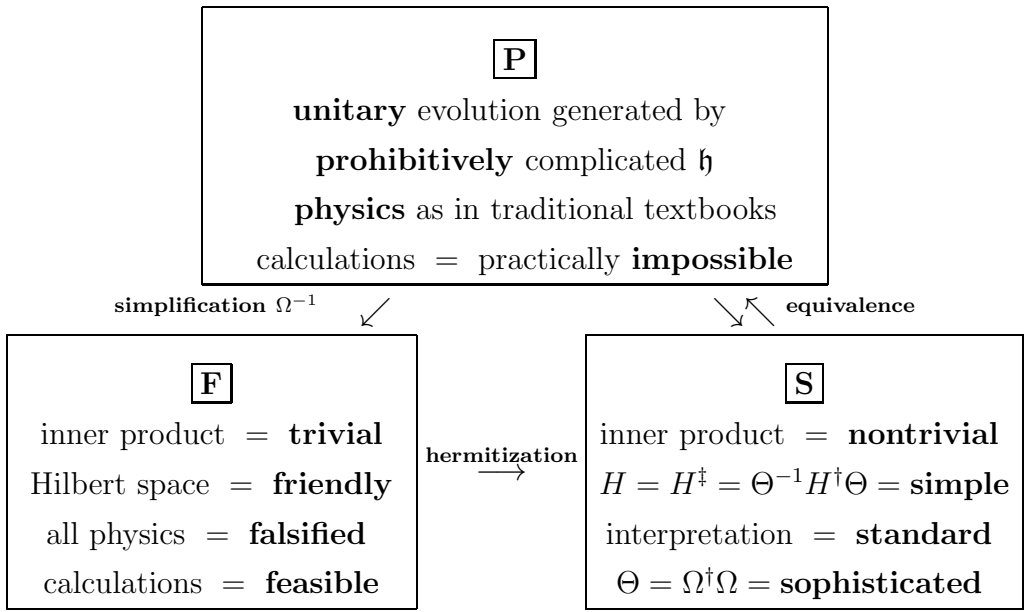
    \begin{array}{|c|}
 \hline
 \vspace{-0.3cm}\\
  {\rm  \fbox{\bf {P}}}\\
 {\rm {\bf unitary}\   evolution\  generated\ by }\ \\
 {\rm  {\bf prohibitively}\ complicated}\
 \mathfrak{h}\\
    \ \ \ \ {\rm  \ {\bf physics}\ as \ in \ traditional \ textbooks } \ \\
  \ \ \ \ {\rm  calculations\ =\  practically\ \bf impossible\ } \ \\
 \hline
 \ea
 \\
 \stackrel{{\bf  simplification}\ \Omega^{-1}}{}
 \ \ \ \
  \swarrow\ \  \  \ \ \ \ \ \
 \ \ \ \ \ \ \ \
  \ \  \  \ \ \ \ \ \
 \ \ \ \ \  \searrow \nwarrow\
 \stackrel{\bf   equivalence}{}\\
 \begin{array}{|c|}
 \hline
 \vspace{-0.35cm}\\
  {\rm  \fbox{\bf {F}}}\\
   {\rm   inner \ product\ = \bf \ trivial }  \\
 {\rm  \ Hilbert\ space\ = \bf  \ friendly\ }\\
  {\rm all \  physics\ =  \bf \ falsified\ }  \\
    \ {\rm {  calculations\ =  \bf \  feasible\  } } \\
  \hline
 \ea
 \stackrel{ {\bf  hermitization}  }{ \longrightarrow }
 \begin{array}{|c|}
 \hline
 \vspace{-0.35cm}\\
  {\rm  \fbox{\bf {S}}}\\
  {\rm  inner\ product\ = \bf \ nontrivial \ }  \\
    H=H^\ddagger=\Theta^{-1}H^\dagger\Theta= {\rm \bf simple }  \\
  {\rm interpretation \ = \ \bf standard\ }
 \\
    \Theta=\Omega^\dagger\Omega=   {\rm \bf sophisticated\ } \\
 \hline
 \ea
\\
\\
\\
\ea
 \een

\section{A note on the history and applications.  \label{jednasecbe}}

\subsection{The birth and the resolution of the puzzle.}

Although the very compact review-like 2002 note  ``The challenge of
non-Hermitian structures in physics'' by A. Ramirez and B. Mielnik
\cite{Bogdan} is merely twelve years old, the subject and its
applications in the various branches of physics developed, in
between, so quickly that one should (and,  in what follows, we are
going to) update some of their conclusions. {\it Pars pro toto},
today, the Ramirez's and Mielnik's citation of the 2001 note
\cite{ptsqw} offering a vague indication of the non-Hermitian
Hamiltonian's having ``link with pseudo-euclidean structures''
\cite{Bogdan} would have to be complemented by the reference to the
subsequent 2004 paper \cite{Batal}. In the latter text the authors
considered the same illustrative non-Hermitian square-well
Hamiltonian but they already were able to explain its full
compatibility with the first principles of conventional quantum
theory. In this manner the latter authors provided a virtually
exhaustive resolution of all of the related apparent paradoxes.
Thus, in brief, one can only repeat that after the year 2004, the
``consistent interpretation'' of non-Hermitian quantum Hamiltonians
$H \neq H^\dagger$ of Bender with Boettcher \cite{BB} could not have
been declared ``missing'' by the authors of Ref.~\cite{Bogdan}
anymore.

\subsection{The current state of art and the continuing emergence of new
puzzles.}


A brief recollection of the developments in the field during the
last twelve years reveals that the related research activities did
not stop after 2004. Naturally, an understanding of the basic idea
was already available but multiple open questions survived. Many of
them were already asked around the end of the millennium, i.e.,
immediately after the publication of the inspiring letter \cite{BB}.
During a few years, many non-Hermitian Hamiltonians $H^{(NH)}$ with
real spectra were then analyzed by many authors. Still, using the
words of {\it loc. cit.}, ``in all of these designs'' the proper
``statistical interpretation [was] still missing'' \cite{Bogdan}.

Fortunately, as we already mentioned, the progress was quick. Around
the year 2004, virtually all of the essential connections between
the exotic-looking $H^{(NH)}$ and the conventional quantum theory
seem to have been already established. Still, new ideas kept
emerging even after the year 2004. Typically, the ambiguity problems
concerning the assignment of a metric $\Theta$ to a given
Hamiltonian $H$ were never completely abandoned. Between the years
2007 - 2010 people also re-opened \cite{Jones} and solved
\cite{scatt} the puzzling unitarity/non-unitarity conflict in the
scattering  arrangement. Similarly, due to the apparent failure of
the semi-classical approximations, a new crisis emerged very
recently \cite{Siegl}. Last though not least, even the lasting
conflict between the intuitive and rigorous quantum-theoretical
perception of the concept of locality did also hit the PTSQM theory
in the past \cite{uwebrach} as well as very recently \cite{Lee}.
Nevertheless, in a way paralleling these fluctuations, various
versions of the general THS theory may be now declared to have
acquired, under several sophisticated technical assumptions, a more
or less closed and final-looking form.

%
%
%
%
%
%
%
%
%

\section{The challenge of time-dependent non-Hermitian Hamiltonians
in Quantum Mechanics.
 \label{dvasec}}

\subsection{A remark on terminology.}

In the next-to-perfect Mostafazadeh's review paper \cite{ali} the
physicists read, with satisfaction, that the ``time-dependent
quasi-Hermitian Hamiltonians arise naturally in the application of
pseudo-Hermitian quantum mechanics in quantum cosmology''. At the
same time the mathematicians could feel puzzled when reading there
that ``in pseudo-Hermitian quantum mechanics we are bound to use
quasi-stationary Hamiltonians'' defined as ``admitting a
time-independent metric''. Puzzling as a comparison of these two
statements may sound (cf. also the unpublished discussion of this
topic in arXiv \cite{webbed}), it in fact merely reflects the
Mostafazadeh's unexplained decision of working, exclusively, with
the {\em observable} generators of the quantum time evolution. In
other words, the scope of Mostafazadeh's PHRQM formulation remains
restricted to the above-mentioned Schr\"{o}dinger
Eqs.~(\ref{Fschrodinger}) in which the generators of evolution $H$
remain compatible with the Dieudonne's quasi-Hermiticity constraint
(\ref{Dieudonne}). Then, together with the standard requirements of
the unitarity of the theory this would really imply that we must
have $\Theta \neq \Theta(t)$, indeed.

In this sense, the THS time-dependent-metric representation of a
quantum system as proposed in Refs.~\cite{SIGMA,timedep} may be
perceived as a further nontrivial generalization of the Bender's
time-independent-metric PTSQM frame as well as of the Mostafazadeh's
time-independent-metric formalism of PHRQM.

\subsection{The challenge of time-dependent metrics.}

Once we admit that the crypto-Hermitian time-evolution
time-independent-metric law (\ref{Fschrodinger}) may be further
generalized, our constructive considerations  become straightforward
(cf. \cite{SIGMA,timedep} for details). First of all, admitting the
explicit time-variability of the Dyson's map $\Omega=\Omega(t)$ of
the ket-vector spaces ${\cal H}^{(F)} \to {\cal H}^{(P)}$, i.e.,
postulating the relation
 \be
 |\psi(t)\rangle^{(P)} =\Omega(t)\,|\psi(t)\rangle^{(F)}
 \ee
an elementary insertion of this ansatz in the original
Schr\"{o}dinger Eq.~(\ref{schrodinger}) immediately yields the
properly modified form of its equivalent representation in the
friendlier Hilbert space ${\cal H}^{(F)}$,
 \begin{equation}
 \label{Sschrodinger}
 i\hbar\frac{\partial}{\partial t}|\psi(t)\rangle^{(F)} =
 G(t)\,|\psi(t)\rangle^{(F)}\,,
 \ \ \ \ \  G(t)=H(t)-\Sigma(t)\,.
 \end{equation}
In this evolution equation the new, time-dependent isospectral image
 \be
 H(t) = \Omega^{-1}(t)\,\mathfrak{h}(t)\,\Omega(t) =H^\ddagger(t)
 = \Theta^{-1}(t)\,H^\dagger(t)\,\Theta(t)\,
 \ee
of the original Hamiltonian enters the time-dependent generator of
quantum evolution in combination with the so called \cite{unitary}
quantum Coriolis force
 \be
 \Sigma(t)={\rm i}\Omega^{-1}(t)\,\dot{\Omega}(t)\,,
 \ \ \ \dot{\Omega}(t)=\partial_t{\Omega}(t)\,.
 \ee
We should add that  the emergence of the Coriolis term $\Sigma(t)$
simply reflects the emergence of the manifest time-dependence of the
inner products in the alternative physical and
nontrivial-metric-endowed third Hilbert space ${\cal H}^{(S)}$.
Secondly, we should emphasize that the latter space still coincides
with  ${\cal H}^{(F)}$ up to the metric, i.e., as a topological
vector space \cite{ali}. Thirdly, we may return now to the first
paragraph of this section and see that separately, {\em both} the
``virtual-force'' Coriolis operator $\Sigma(t)$ {\em and} the
related time-dependent generator  $G(t)=H(t)-\Sigma(t)$ become, in
general, {\em unobservable}. In other words, the requirement of the
manifest time-dependence of the generator in our most general but
still {\em unitarity-guaranteeing} Schr\"{o}dinger
Eq.~(\ref{Sschrodinger}) implies that the spectrum of such an
operator $G(t)$ {\em may cease to be real}.

In this sense, the standard textbook Quantum Theory admits its
phenomenologically most general but still mathematically fully
consistent THS representation as introduced in Ref.~\cite{timedep},
reviewed in Ref.~\cite{SIGMA} and described by the following amended
diagram
 \ben
  \vspace{-1cm}
  \ba
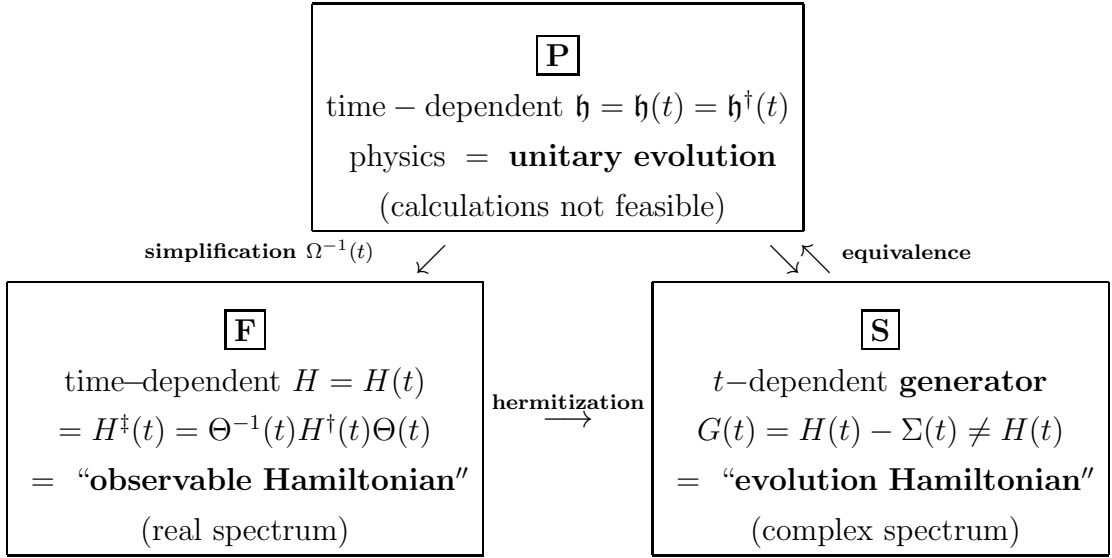
    \begin{array}{|c|}
 \hline
 \vspace{-0.3cm}\\
  {\rm  \fbox{\bf {P}}}\\
 {\rm  { time-dependent}}\
 \mathfrak{h}=\mathfrak{h}(t)=\mathfrak{h}^\dagger(t)\\
     \ {\rm  \ { physics}\ = \  \bf unitary\   evolution } \ \\
  \ {\rm  (calculations\    not\ feasible)\ } \ \\
 \hline
 \ea
 \\
 \stackrel{{\bf  simplification}\ \Omega^{-1}(t)}{}
 \ \ \ \
  \swarrow\ \  \  \ \ \ \ \ \
 \ \ \ \ \ \ \ \
  \ \  \  \ \ \ \ \ \
 \ \ \ \ \  \searrow \nwarrow\
 \stackrel{\bf   equivalence}{}\\
 \begin{array}{|c|}
 \hline
 \vspace{-0.35cm}\\
  {\rm  \fbox{\bf {F}}}\\
    {\rm  { time\!\!-\!\!dependent}}\
  H=H(t)\\
    =H^\ddagger(t)=\Theta^{-1}(t)H^\dagger(t)\Theta(t)  \\
   {\rm   \ = \ \bf ``observable\ Hamiltonian''}\\
   \ {\rm { (real\ spectrum) \ } } \\
  \hline
 \ea
 \stackrel{ {\bf  hermitization}  }{ \longrightarrow }
 \begin{array}{|c|}
 \hline
 \vspace{-0.35cm}\\
  {\rm  \fbox{\bf {S}}}\\
   t{\rm  -dependent \ } {\rm \bf  generator}  \\
  G(t)=H(t)-\Sigma(t) \neq H(t)\\
   {\rm   \ = \ \bf ``evolution\ Hamiltonian''}\\
   \ {\rm { (complex\ spectrum)  } } \\
 \hline
 \ea
\\
\\
\\
 \ea
 \een

\section{Properties of sophisticated physical
Hilbert space ${\cal H}^{(S)}$.
 \label{dvasecbe}}

\subsection{A mixed blessing of the use of the time-dependent Dyson maps.}

The two-step realization $P \to F \to S$ of the unitary equivalence
between the two alternative {\em physical} Hilbert spaces ${\cal
H}^{(P)}$ and ${\cal H}^{(S)}$ has its merits (e.g., a
simplification $\mathfrak{h}(t) \to H(t)$ of the observable of the
instantaneous but still, in principle, measurable $P-$space-based
energy of the system) and shortcomings (e.g., the use of a
frequently rather misleading terminology). Still, the above-cited
emergence of the non-Hermitian plus time-dependent forms of the
hiddenly unitary quantum evolution law (\ref{Sschrodinger}) ``in
\ldots quantum cosmology'' \cite{ali} offers a sufficiently
persuasive motivation for the mathematical study as well as for new
proposals of phenomenological applications of the THS
representations of quantum systems which are made more flexible by
the permission of time-dependence in the underlying Dyson's maps
$\Omega(t)$.

Naturally, the price to be paid for the maximally enhanced
flexibility of Eq.~(\ref{Sschrodinger}) is not too low. In
particular, the original motivation of the formalism (which proved
fairly persuasive in theory, plus strong in applications) gets
perceivably weakened in the time-dependent THS case \cite{unitary}.
Moreover, the technical difficulties further increase if we decide
to invert the original ``Dyson's'' direction $P \to F \to S$ of the
construction as incorporated in the PHRQM formalism of
Refs.~\cite{ali,Dyson,Geyer} in which one started all considerations
from a given pair of operators $\mathfrak{h}\neq \mathfrak{h}(t)$
and $\Omega\neq \Omega(t)$. Indeed, even in the perceivably more
restrictive but still time-independent-metric-unsing PTSQM formalism
as summarized in Ref.~\cite{Carl} the situation appeared complicated
since the pair of operators $\mathfrak{h}\neq \mathfrak{h}(t)$ and
$\Omega\neq \Omega(t)$ only had to be reconstructed at the very end
of all of the constructive manipulations (cf., e.g., an exactly
solvable model \cite{allmet} for illustration).

All this explains why the current progress in the cosmological
applications of the THS formalism (cf., e.g., their first
preliminary samples in \cite{Bang}) still remains so deplorably
slow. At the same time, the current tradition of the use of just
finite-dimensional Hilbert spaces in the context of quantum control
seems to open new perspectives for applications of the
time-dependent-THS Schr\"{o}dinger Eq.~(\ref{Sschrodinger}). Let us,
therefore, complement our preceding introductory THS outline by a
few most relevant further technicalities.

\subsection{{\em Ad hoc} notation conventions.}

First of all, let us remind the readers of our review
paper~\cite{SIGMA} that in parallel to the above-mentioned formal
coincidence of kets $|\psi(t)\rangle^{(F)}
=|\psi(t)\rangle^{(F)}=|\psi(t)\rangle $ (i.e., to the formal
coincidence of the two ket-vector spaces ${\cal H}^{(F)}$ and ${\cal
H}^{(S)}$) one has to keep in mind that the respective conjugate
dual-space elements {\em alias} primed-vector-space elements {\em
alias} linear functionals (i.e., in the Dirac's terminology, the
bra-vectors $^{(F)}\!\langle \psi| \in \left [ {\cal H}^{(F)}\right
]'$ and $\,^{(S)}\!\langle \psi| \in \left [ {\cal H}^{(S)}\right
]'$ which are assigned to the corresponding ket vectors via the
respective Hermitian-conjugation antilinear operations ${\cal T}$)
remain different,
 \be
 ^{(S)}\!\langle \psi|\,\equiv\ ^{(F)}\!\langle \psi|\,\Theta
 \,\neq \  ^{(F)}\!\langle \psi|\,.
 \ee
In order to emphasize this important feature of the $F
\leftrightarrow S$ corrrespondence we shall use the notation of
\cite{SIGMA} and abbreviate $\langle\!\langle \psi| \,\equiv\,
\,^{(S)}\!\langle \psi| \in \left [ {\cal H}^{(S)}\right ]'$ in what
follows.

In order to suppress confusion we shall also accept another
convention that all of the eligible Hermitian conjugations ${\cal
T}: |\bullet \kt \to \langle \bullet |$ will occur without
superscripts, i.e., they will always be understood as performed
solely in the trivial-metric spaces, i.e., just in our $^P-$ or
$^F-$superscripted Hilbert spaces. This means that in our present
paper we shall never employ the abbreviated and metric-dependent
Hermitian-conjugation operation ${\cal T}^{(S)}$. Thus, for example,
the inverse conjugation ${\cal T}^{-1}: \langle\!\langle \bullet|
\to |\bullet\kkt$ will be always understood as performed just in the
friendly, $^F-$superscripted Hilbert space, etc.

The use of such notation conventions enables us to characterize the
unitary equivalence between our $^P-$ and $^S-$superscripted Hilbert
space in an extremely compact manner, viz., via the following
coincidence of the respective inner products,
 \be
 \pbr \psi_1|\psi_2 \pkt \ (\,\equiv
  \,\,^{(P)}\langle  \psi_1|\psi_2 \kt^{(P)}\,)\
 = \langle\!\langle  \psi_1|\psi_2 \kt
  \ (\,\equiv \,\,^{(S)}\langle  \psi_1|\psi_2 \kt^{(S)}\,)\,.
 \ee
Moreover, the conventional textbook use of an orthonormalized basis
$\{ \,|n\pkt\,\}$ in ${\cal H}^{(P)}$ may be immediately paralleled
by its $^S-$superscripted-Hilbert-space (bi)orthonormal-basis
descendant with the respective kets $|n\kt$ and bras
$\langle\!\langle n|$, etc. On these grounds one characterizes a
{\em physical} state of a given quantum system {\em either} by the
kets $|\psi(t)\pkt$ and bras $\pbr \psi(t)|$ in the
$^P-$superscripted representation {\em or, alternatively}, by the
``simpler'' kets $|\psi(t)\kt$ and bras $\langle\!\langle \psi(t)|$
in their preferable but formally strictly unitarily equivalent
$^S-$superscripted representation.

\subsection{Evolution control by {\em two}
 Schr\"{o}dinger equations.}

In the constructive mathematical perspective a decisive
THS-representation advantage is that one never has to leave the
auxiliary friendly space, treating the structure-reflecting concepts
and symbols like, e.g., $\langle\!\langle \psi(t)|$ or $H^\ddagger$
as the mere metric-containing abbreviations. Moreover, as we already
mentioned, a key benefit of our conventions is that after an
ultimate return to the friendly Hilbert space we have got rid of all
of the superscripts. In particular, the fully general non-Hermitian
THS quantum evolution process as described in Ref.~\cite{SIGMA} may
be now perceived as initiated, at time $t=0$ and in its friendly
${\cal H}^{(F)}$ representation, by the choice of {\em two} initial
ket-vectors $|\psi(0)\kt$ and  $|\psi(0)\kkt$, with the latter one
being formally expressible, in the cases when we know the metric, as
the metric-multiple $\Theta(0)\,|\psi(0)\kt$. Next, in the
Dyson-inspired direct $P \to F \to S$ recipe one has to know the
generator $G(t)$ and, as long as $G(t) \neq G^\dagger(t)$, one must
solve the {\em two} time-evolution Schr\"{o}dinger equations,
 \be
 \label{Skschrodinger}
 i\hbar\frac{\partial}{\partial t}|\psi(t)\kt =
 G(t)\,|\psi(t)\kt\,
  \ee
 \be
 \label{Sbschrodinger}
 i\hbar\frac{\partial}{\partial t}|\psi(t)\kkt =
 G^\dagger(t)\,|\psi(t)\kkt\,
  \ee
(incidentally, notice an unfortunate misprint in \cite{SIGMA}). One
can also find another benefit of our notation in the subsequent
elementary re-derivation of formula
 \be
 \partial_t\,\langle\!\langle \psi(t)|\psi(t)\kt =0\,,
 \ee
i.e., in a reconfirmation of conservation law for the norm of state
$\psi(t)$ when considered in its amended physical
$\,^{(S)}-$superscripted representation.

\section{Time-dependent Dyson maps in quantum control setup.
\label{trisec}}

\subsection{A sample of the realization of the project of
generalized non-Hermitian quantum control.}

In the THS generalization the traditional QC superposition ansatz
(\ref{postulate}) may be made less restrictive in several directions
involving, first of all, several alternative real-control-function
assumptions. Thus, the traditional Hermitian QC-related
postulate~(\ref{postulate}) may be replaced, say, by its analogues
describing the ``evolution Hamiltonian'' with a complex spectrum
 \be
 G(t)=G_{0} +
 \sum_{k=1}^{K_G}u_{k}(t)G_{k}\,
 \label{Gpostulate}
 \ee
and/or the ``observable Hamiltonian'' with the (time-varying but, in
principle, measurable) instantaneous-energy real spectrum,
 \be
 H(t)=H_{0} +
 \sum_{m=1}^{K_H}z_{m}(t)H_{m}\,
 \label{Hpostulate}
 \ee
etc. Naturally, the most fundamental innovation may be expected to
result from the highly nontrivial nature of the non-unitary,
manifestly time-dependent Dyson's maps, say, of the same multinomial
form
 \be
 \Omega(t)=\Omega_{0} +
 \sum_{n=1}^{K_\Omega}v_{n}(t)\Omega_{n}\,.
 \label{Opostulate}
 \ee
Obviously, as long as the knowledge of $\Omega(t)$ implies the
knowledge of the Coriolis term $\Sigma(t)$, the role of assumption
(\ref{Opostulate}) seems fundamental. Only when we choose
$K_\Omega=1$ and set $\Omega_0=0$ we still obtain a transparent
multinomial-operator toy model with metric
$\Theta(t)=v^2(t)\Theta_1$ and with a diagonal-matrix Coriolis
operator $\Sigma(t)={\rm i}\dot{v}(t)/v(t)\,I=-{\rm i}w(t)\,I$.

\section{The ultimate reconstruction challenge. \label{point}}

In the context of our preceding illustrative example we may prolong
our methodical analysis and choose, say, $K_H=1$. This will enable
us to insert all ansatzs in the Dieudonn\'{e}'s observability
requirement (\ref{Dieudonne}). With the real control function
$z_1(t)=z(t)$, this requirement becomes time-variation-independent
and it may be separated and solved elementwise, yielding two
conditions
 \be
 H_0^\dagger\,\Theta_1=\Theta_1\,H_0\,,
 \ \ \ \ \ \ \
 H_1^\dagger\,\Theta_1=\Theta_1\,H_1\,.
 \ee
If the solution $\Theta_1$ exists we shall be already able to derive
the closed form of the generator
 \be
 G(t)=H_0+u(t)\,H_1+w(t)\,H_2
 \ee
with $K_G=2$ and $ H_2={\rm i}\,I\,$. Thus, in a way, we shall
return to a more or less standard QC scenario, with the main
difference and innovation resulting from our new freedom of having
the generator $G(t)$ which is non-Hermitian and which is even
non-quasi-Hermitian (i.e., which does have complex eigenvalues).

In the latter context let us finally recall an unpublished preprint
\cite{Bila} in which Hynek B\'{\i}la tried to study a few more
concrete non-Hermitian toy models living in the two-dimensional,
i.e., in the first nontrivial friendly complex Hilbert space ${\cal
H}^{(F)} = \ell^{(2)}$. This study revealed that one could also
avoid the reference to the THS Schr\"{o}dinger equations completely.
In such an approach the necessary time-dependent metric $\Theta(t)$
has to be reconstructed via direct solution of the corresponding
operator evolution differential equation of the
Heisenberg-representation-resembling form which follows immediately
from the definition of $\Sigma(t)$,
 \be
 {\rm i}\partial_t\Theta(t)=
 G^\dagger(t)\Theta(t)-\Theta(t)\,G(t)\,.
 \label{paracau}
 \ee
Unfortunately, the B\'{\i}la's preliminary results were never
completed (cf. also \cite{Bilabe,Bilance}). Perhaps, the project
itself could still acquire a new life in the non-Hermitian QC
context.

In the conclusion let us add that the metric-determining ``parallel
Cauchy problem'' (\ref{paracau}) could be addressed by various
techniques and under a multitude of approximations but in the QC
context, the use of a finite-dimensional Hilbert-space approximation
seems most promising, especially because it parallels the common
practice used in the standard Hermitian models \cite{Dong}. Thus, we
believe that in the nearest future, the traditional Hermiticity
condition $H(t)=H^\dagger(t)$ need not remain obligatory and
uncircumventable, anymore.

\end{document}